\documentclass[a4paper,11pt]{article}
\pdfoutput=1 % if your are submitting a pdflatex (i.e. if you have
\usepackage{jinstpub} % for details on the use of the package, please
\title{\boldmath First Annealing Studies of Irradiated Silicon Sensors with Modified ATLAS Pixel Implantations}

\author[a,1]{M. Wagner,\note{Corresponding author.}}
\author[a]{A.Gisen,}
\author[a]{M. H\"otting,}
\author[a]{V. Hohm,}
\author[a]{C. Krause,}
\author[a]{K. Kr\"oninger,}
\author[a]{A. Kroner,}
\author[a]{\mbox{J. L\"onker},}
\author[a]{M. Muschak,}
\author[a]{J. Weingarten}
\author[a]{and F. Wizemann}

\affiliation[a]{Experimentelle Physik IV, TU Dortmund,\\Otto-Hahn-Str. 4a, 44227 Dortmund, Germany}
\emailAdd{mareike.wagner@tu-dortmund.de}

\abstract{ Planar silicon pixel sensors with modified n$^+$-implantation shapes based on the IBL pixel sensor were designed in Dortmund. The sensors with a pixel size of $250\,$\textmu m $\times$ $50\,$\textmu m are produced in n$^+$-in-n sensor technology.

The charge collection efficiency should improve with electrical field strength maxima created by the different n$^+$-implantation shapes. Therefore, higher particle detection efficiencies at lower bias voltages could be achieved. The modified pixel designs and the IBL standard design are placed on one sensor to test and compare the designs. The sensor can be read out with the FE-I4 readout chip.

At the iWoRiD 2018, measurements of sensors irradiated with protons and neutrons respectively at different facilities were presented and showed incongruent results. Unintended annealing during irradiation was considered as an explanation for the observed differences in the hit detection efficiency for two neutron irradiated sensors.
This hypothesis will be examined and confirmed in this work, presenting first annealing studies of sensors irradiated with neutrons in Ljubljana. 
}

\keywords{Hybrid detectors, Particle tracking detectors, Particle tracking detectors (Solid-state detectors), Solid state detectors, Radiation-hard detectors, Radiation damage to detector materials (solid state)}

\arxivnumber{1908.10652} % only if you have one

\proceeding{21$^{\text{st}}$ International Workshop on Radiation Imaging Detectors\\
7-12 July 2019\\
Kolympari, Chania, Crete, Greece}

\begin{document}
\maketitle
\flushbottom

\section{Introduction}
The ATLAS experiment \cite{ATLAS} at CERN was equipped with the Insertable B-Layer (IBL) \cite{IBL} between the existing inner pixel layer and a new beam pipe during the long shutdown one to improve the tracking performance. The planar and 3D sensors of the IBL are exposed to a high flux of ionizing radiation because of the close position to the interaction point. The planar sensors are designed to withstand a fluence of $5\cdot 10^{15}\,\text{n}_\text{eq}/\text{cm}^{2}$. They are produced in n$^+$-in-n sensor technology and have a pixel pitch of $250\,$\textmu m $\times$ $50\,$\textmu m \cite{sensor1, sensor2}. The 80 columns and 336 rows of pixels are read out with the Front-End-I4 readout chip \cite{FE-I4}.

While operating the detector, the radiation damage of the sensors will increase their leakage current and depletion voltage, whereas the charge collection efficiency decreases. In order to continue to obtain a high particle detection efficiency, the bias voltage needs to be increased gradually. Voltages up to $1000\,$V can be applied to the sensors causing an increase in power consumption. The higher power consumption produces heat which needs to be dissipated by the detectors cooling system to prevent thermal runaway of the sensors' leakage current. Therefore, a higher detection efficiency at lower bias voltage is desirable.

New pixel implantation shapes were designed in Dortmund to achieve electrical field strength maxima in the pixel and thus increase charge collection and particle detection efficiency at lower voltages after irradiation \cite{REINER}.

In the proceeding \emph{Lab and test beam results of irradiated silicon sensors with modified {ATLAS} pixel implantations} \cite{meinPaper}, results of these proton and neutron irradiated modules were presented and showed incongruent results. This paper now examines the hypothesis that an annealing process caused the observed differences.

\section{Design of the Pixel Cell}
\begin{figure}[htbp]
\centering
\includegraphics[width=.6\textwidth]{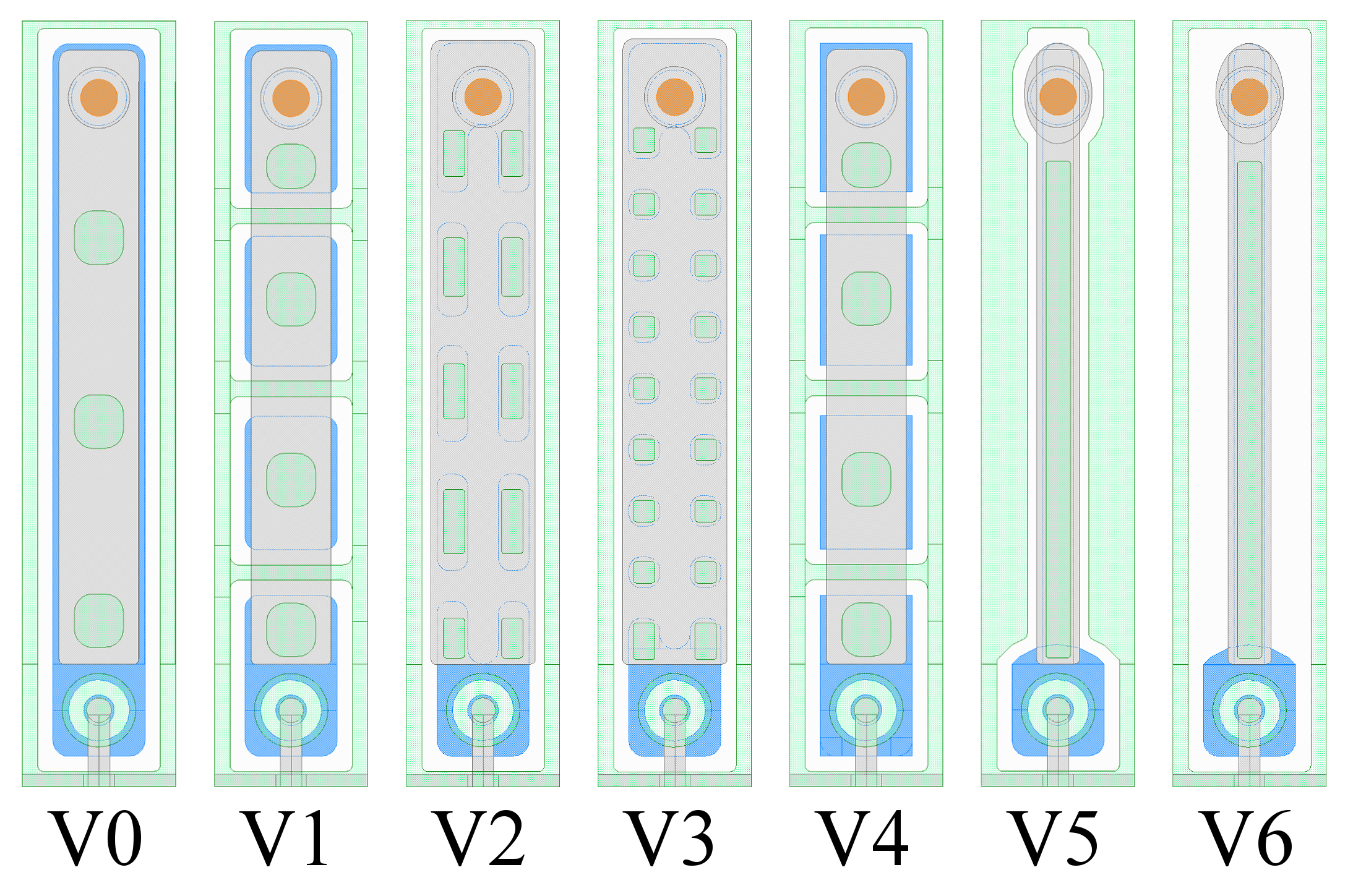}
\caption{\label{fig:design} Schematic drawing of the pixel structure and the shape of its components. The $\text{n}^+$-implantation is shown in blue, the metalization in grey. The areas with high p-spray dose correspond to the nitride-openings as indicated in green. The bump bond pad is depicted in orange \cite{meinPaper}.}
\end{figure}

The baseline for the new designs is the IBL pixel design (see figure \ref{fig:design}, label V0). In the \mbox{$250\,$\textmu m $\times$ $50$ \textmu m} pixel cell, the standard n$^+$-implantation is located centrally with rounded corners to create a homogeneous electrical field in the pixel. Moderated p-spray is applied to isolate neighboring pixel cells. In the upper region of the pixel cell the bump bond pad is visible which is the connection to the readout chip. The bias dot with connection to the bias grid is positioned at the other end.

Different n$^+$-implantation shapes are realized in the REINER$^2$\note[2]{\textbf{RE}designed, \textbf{IN}novativ, \textbf{E}xciting and \textbf{R}ecognizable} pixel designs V1 to V6 (see figure \ref{fig:design}): For the pixel designs V1 and V4, the n$^+$-implantation is divided in four uniform segments isolated with p-spray. The corners of the n$^+$- implantation of pixel design V1 are rounded, while they are rectangular for pixel design V4. Further division of the n$^+$-implantation are used for pixel designs V2 and V3. No isolation between the segments is used here due to reduced space. A narrowed shape of the metal layer and the n$^+$-implantation are realized in the pixel designs V5 and V6. While in V6 the region for moderated p-spray is the same as for the standard pixel, the section is increased to the inner part of the pixel cell for design V5.

The connection to the bias dot and bias grid is the same for all pixel designs and causes a loss in efficiency as mentioned in \cite{effVerlust1, effVerlust2}. This will not be taken into account in this proceeding as it focuses on the influence of the different pixel shapes. 

\section{Sensors and Modules}
The REINER pixel sensor permits measurement of all the described pixel types on one sensor. It is produced in n$^+$-in-n sensor technology with a bulk thickness of $200\,$\textmu m. The pixel matrix of 80 columns and 336 rows is read out with the FE-I4 readout chip via bump bonds.

\begin{figure}[htbp]
\centering
\includegraphics[width=1.0\textwidth]{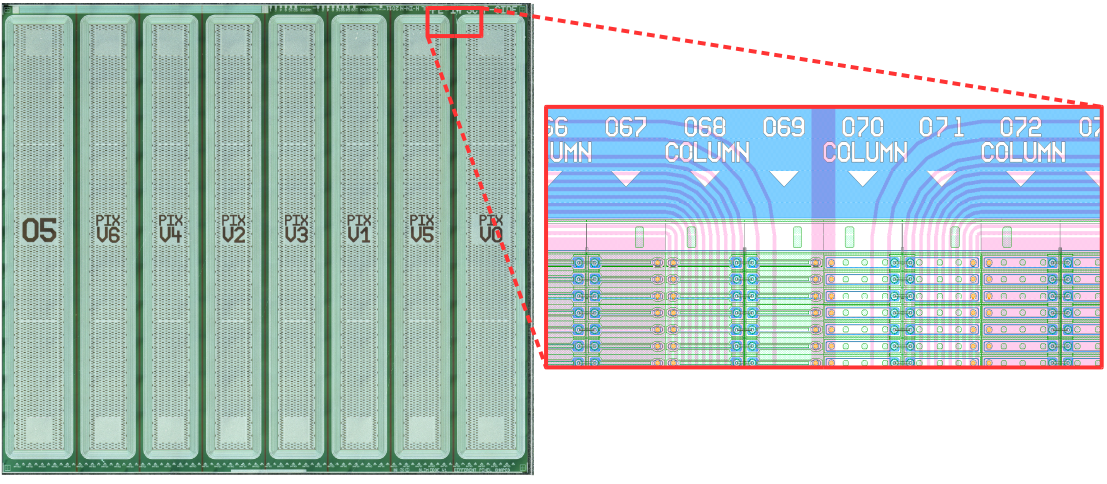}
%\qquad
%\includegraphics[width=.4\textwidth]{Bilder/GuardRinge.png}
\caption{\label{fig:sensor} Left: P-side of a REINER pixel sensor with six modified structures and two IBL standard structures (05 and V0). Right: Magnification of the area between two structures. The second to last and last column of each structure are beneath guard rings \cite{meinPaper}.}
\end{figure}

The sensor is divided in eight structures consisting of ten columns and 336 rows of the same pixel design. The outermost structures on the sensor comprise the standard IBL pixel design (labeled as 05 and V0). In between these standard structures, structures with modified pixel designs (V1 to V6) are placed (see figure \ref{fig:sensor}, left). Each structure has its own p$^+$-implant and two HV pads. All structures are surrounded by thirteen guard rings beneath the second to last pixel column and the last pixel column (see figure \ref{fig:sensor}, right). In this way, measurements of the structures independent from each other are possible.

To compare the performance of the different pixel designs, several sensors are bump bonded to FE-I4 readout chips and tested before and after irradiation in irradiation facilities. The results presented in this work are obtained with modules irradiated with neutrons at the TRIGA reactor in Ljubljana \cite{Ljubljana} and at the Sandia Annular Core Research Reactor$^4$\note[4]{https://www.sandia.gov/research/facilities/annular\_core\_research\_reactor.html}. They are irradiated with neutrons to target fluences of $1\cdot10^{15}\,\text{n}_\text{eq}/\text{cm}^2$ and $5\cdot10^{15}\,\text{n}_\text{eq}/\text{cm}^2$.

\section{Test Beam Measurements}
The sensors are investigated in test beam measurements at the SPS-Beam Line H6 at CERN with a pion beam energy of $120\,$GeV and at the beam line 22 at DESY \cite{DESYPaper} with an electron beam energy of $5\,\text{GeV}$. With the EUDET-type beam telescopes ACONITE at CERN and DURANTA at DESY the hit detection efficiency can be measured with a spatial resolution of less than $5\,$\textmu m (CERN) and $10\,$\textmu m (DESY) \cite{resolution}. A telescope consists of two arms, each equipped with three Mimosa 26 modules \cite{mimosa26}. In between the two arms, a cooled box holds the Devices Under Test (DUTs). The box is flushed with nitrogen and cooled using a chiller (CERN) or dry ice (DESY). A non-irradiated IBL-type module is used as a timing reference sensor.

With this setup the REINER modules are investigated at different positions, bias voltages and tunings of the readout chip. For irradiated REINER modules high efficiencies around $97\,\%$ can be reached for all pixel designs with a high bias voltage. Therefore, operation with a lower bias voltage where differences in the performance of the pixel designs occur is desirable.

The traversing particles produce hits in the telescope modules and the DUTs which are reconstructed to tracks using the software EUTelescope$^5$\note[5]{http://eutelescope.web.cern.ch/}. To get unbiased tracks the hit information of the DUTs is not used for track fitting. About $200\,\text{k}$ to $500\,\text{k}$ events are collected in one "run".

The hit detection efficiency, as one part of the analysis of the performance of the DUTs, is computed with the software tool TBMon2$^6$\note[6]{https://gitlab.cern.ch/tbmon2}.

All reconstructed particles measured by the reference sensor and not in a masked, edge or noisy region of the investigated sensor are used for efficiency calculation and called "tracks". If a "track" is also measured by the investigated sensor, it is defined as a "hit". The hit detection efficiency of a sensor $\epsilon$ is calculated by the quotient of "hits" and "tracks":

\begin{equation}
\epsilon = \frac{n_\text{hits}}{n_\text{tracks}}, \quad \sigma_\epsilon = \sqrt{\frac{\epsilon \cdot (1-\epsilon)}{n_\text{tracks}}}
\end{equation}

To summarize the efficiencies for runs taken under the same conditions the mean efficiency weighted with the number of tracks is calculated. The fluctuation of the efficiency from run to run is determined with the Clopper-Pearson confidence interval \cite{clopper-pearson} with a confidence level of $\gamma=95\,\%$. For the lower interval limit the $\frac{1-\gamma}{2}$-quantile and for the upper limit the $\frac{1+\gamma}{2}$-quantile are calculated \cite{Andreas}.

For the following section the efficiency is calculated by using only the four innermost pixel columns of every structure to neglect the influence of guard rings.

\section{Previous Results}
\begin{figure}[htbp]
\centering
\includegraphics[width=.48\textwidth]{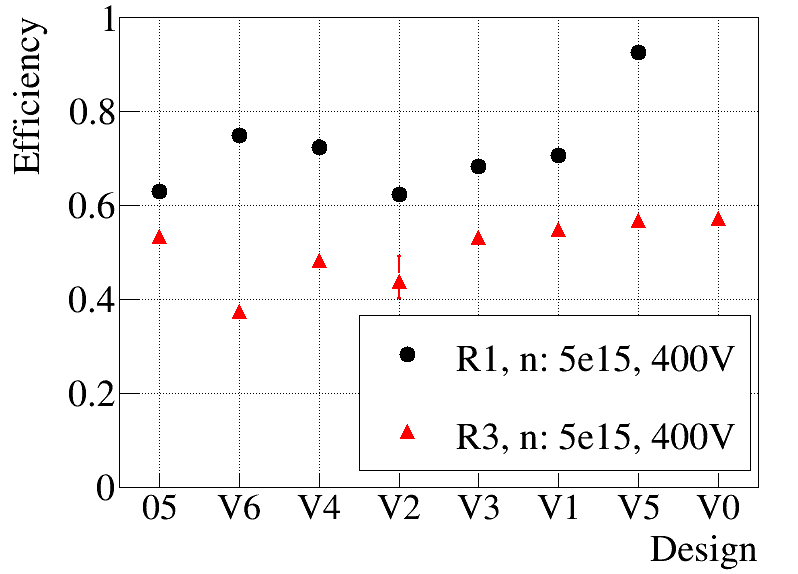}
\includegraphics[width=.48\textwidth]{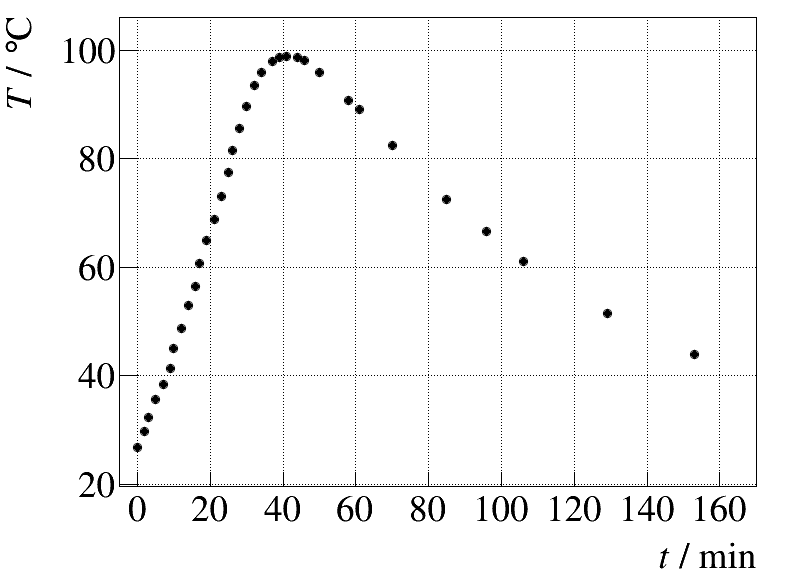}
\caption{\label{fig:bild1} Left: Efficiencies of the different pixel designs for two sensors irradiated with neutrons to a target fluence of $5\cdot 10^{15}\,\text{n}_\text{eq}/\text{cm}^2$ at $400\,$V (Tuning: $3200\,$e threshold, $6\,$ToT at $20\,$ke). R1 is irradiated at Sandia while R3 is irradiated in Ljubljana. Right: Temperatur profile during the neutron irradiation of the module R1 at Sandia.}
\end{figure}

The measurements presented in \cite{meinPaper} showed that for non-irradiated modules the hit detection efficiency for the different pixel designs is consistent. But for modules irradiated with neutrons or protons the hit detection efficiencies are not consistent. Even the two neutron irradiated sensors R1, irradiated at Sandia, and R3, irradiated in Ljubljana, to the same target fluence of $5\cdot 10^{15}\,\text{n}_\text{eq}/\text{cm}^2$ and measured at the same voltage and tuning ($3200\,$e threshold and a ToT response of $6\,$ at a reference charge of $20\,$ke) show dissimilar results (see figure \ref{fig:bild1}, left). For the module R1, irradiated at Sandia, the pixel designs with narrowed n$^+$-implantation V5 and V6 reach the highest efficiencies at $400\,$V. No efficiency of the pixel design V0 was measured for module R1 as the beam spot was focused on the left side of the module and did not cover pixel design V0.

For the Ljubljana irradiated module R3, the pixel design with narrowed n$^+$-implantation V6 has the lowest efficiency while all other pixel designs have similar efficiencies of approx. $50\,\%$.

These diverging results might have been caused by the different neutron energy spectra of the two irradiation facilities while another hypothesis is a different temperature of the modules during irradiation leading to annealing of defects. In Ljubljana the maximum temperature of the modules was $45^\circ$C while they reached about $100^\circ$C in the irradiation at Sandia (see figure \ref{fig:bild1}, right).

To test the annealing hypothesis, the modules R3 and R9 were irradiated with neutrons in Ljubljana to different target fluences and were now annealed in several steps at $80^\circ$C.

\section{First Annealing Results}
For the annealing procedure the climate chamber is preheated to $80^\circ$C before the module is inserted. The temperature of the module is monitored and it is not biased during the annealing. The cooling down afterwards takes place at room temperature and the module is put back in the freezer for storage. After each annealing step IV scans and test beam measurements are performed at DESY.
 
\begin{figure}[htbp]
\centering
\includegraphics[width=.7\textwidth]{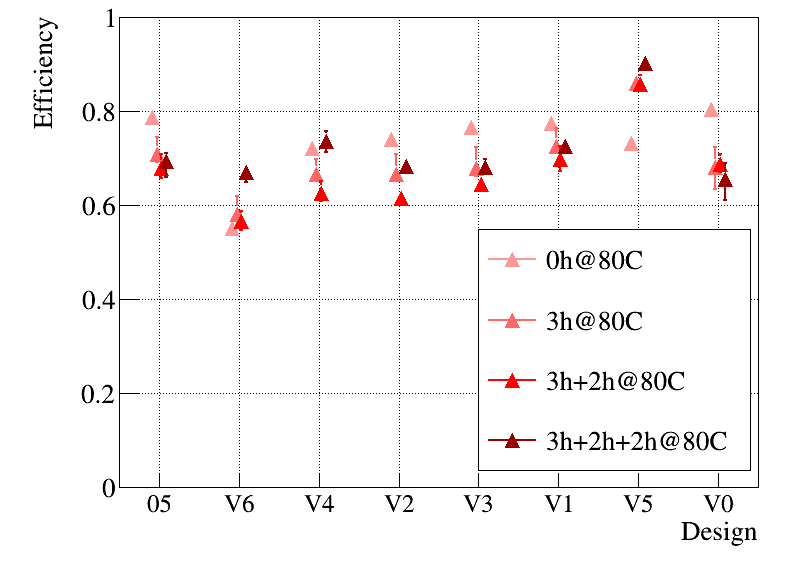}
\caption{\label{fig:bild2} %Left: Annealing profile for the first annealing step for three hours at $80^\circ$C of the module R3. Right: 
Efficiencies of the different pixel designs for the sensor R3 after various annealing steps at $300\,$V (Tuning: $1600\,$e threshold, $6\,$ToT at $20\,$ke).}
\end{figure}

To determine if differences in the efficiency occur with annealing at all, the intention was to anneal a module for a long time. Therefore, the first annealing step of module R3 was three hours at $80^\circ$C. The following annealing steps lasted for two hours each. %The sensor temperature profile for the first annealing step of module R3 for three hours at $80^\circ$C is shown on the left in figure \ref{fig:bild2}. The following annealing steps of the module R3 lasted for two hours each. After each annealing step IV scans and test beam measurements at DESY are performed.\\
The results of the hit detection efficiency for the different pixel designs for module R3 after several annealing steps are presented in figure \ref{fig:bild2}. For these measurements at $300\,$V the module was tuned to a threshold of $1600\,$e and a ToT response of $6\,$ at a reference charge of $20\,$ke. The standard pixel designs 05 and V0 reach the highest efficiencies for the non-annealed case while the pixel designs with narrowed n$^+$-implantation V5 and V6 are less efficient compared to all other designs. This behavior appears to be independent on the threshold as figure \ref{fig:bild1} on the left shows where the sensor was measured at a threshold of $3200\,$e while in figure \ref{fig:bild2} the module was tuned to a threshold of $1600\,$e.

With the first annealing step of three hours the efficiency drops for all pixel designs except for the designs with a narrowed n$^+$-implantation. Compared to the standard pixel design, the hit detection efficiency drops by $8\,\%$ while for pixel design V5 the efficiency increases by more than $12\,\%$. After this first annealing step, the efficiency of pixel design V5 is even higher than the non-annealed results of the standard designs. With further annealing steps the efficiency of the standard pixel designs is almost constant, but for pixel designs V5 and V6 where the efficiency increases.

\newpage

\begin{figure}[htbp]
\centering
\includegraphics[width=.7\textwidth]{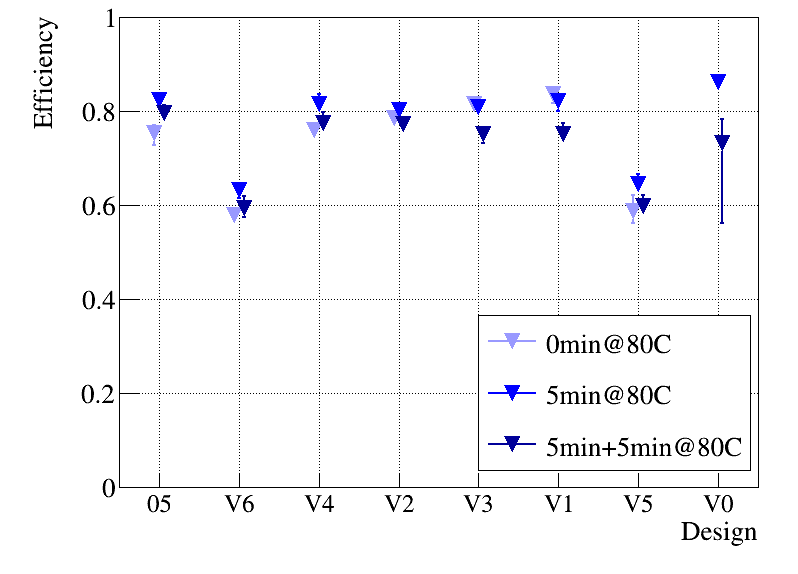}
\caption{\label{fig:bild3} Efficiencies of the different pixel designs for the sensor R9 after various annealing steps at $100\,$V (Tuning: $1600\,$e threshold, $6\,$ToT at $20\,$ke).}
\end{figure}

These results confirm the hypothesis that annealing caused the narrowed pixel designs to be more efficient compared to the standard pixel design. To determine if this effect is dependent on the particle fluence and to see when the effect becomes relevant, the module R9 was irradiated with neutrons in Ljubljana to a target fluence of $1\cdot10^{15}\,\text{n}_\text{eq}/\text{cm}^2$ and measured with a shorter annealing step time of five minutes at $80^\circ$C. The results of the hit detection efficiency for the module R9 at $100\,$V, tuned to a theshold of $1600\,$e and a ToT response of $6\,$ at a reference charge of $20\,$ke, are presented in \mbox{figure \ref{fig:bild3}}.

For the non-annealed measurement the pixel designs with narrowed n$^+$-implantation have the lowest efficiency. Because the beam spot is focused on the left part of the sensor, there is no data point for the pixel design V0. After the first annealing step an increase in efficiency is visible for the pixel designs with a narrowed n$^+$-implantation and the standard pixel design. However, after the second annealing step the efficiency decreases for all designs. Compared to the non-annealed results, the efficiency after the second annealing step is still higher for the standard pixel design and the designs with narrowed n$^+$-implantation.

It is known, that annealing performed in a short time scale is able to reduce the effective doping concentration in the sensor material and more signal at the same voltage can be measured. This process is called beneficial annealing. In contrast, reverse annealing degrades sensor properties in the long term. For more information on annealing effects see \cite{Moll}. Beneficial annealing can explain the higher efficiencies after the first annealing step of the module R9 for the shorter annealing time. But with the second step the reverse annealing becomes dominant and the efficiency drops for all designs. In case of the module R3 only the reverse annealing can be observed because of the long term annealing of three hours at $80^\circ$C.

The observed changes of the hit detection efficiency with annealing are caused by at least two different effects. With annealing the depletion voltage and the charge collection efficiency change. Thus, if one of these effects depend on the pixel design, different hit detection efficiencies might be observed. Also charge multiplication could explain the higher efficiencies, which was also observed in \cite{Milovanoivi,Kramberger1}, where neutron irradiated n-in-p micro-strip sensors showed higher electrical fields and therefore charge multiplication close to the strips.

\section{Summary and Outlook}
The results of the Ljubljana neutron irradiated module R3 indicate that long term annealing caused a higher hit detection efficiency for the pixel designs with narrowed n$^+$-implantation compared to all other designs. As a result of this, the higher detection efficiency observed for pixel designs with narrowed n$^+$-implantation on the neutron irradiated sensor R1 at Sandia (see \cite{meinPaper}) are explained as being due to the higher temperature during the irradiation.

As this effect was not visible for the module R9, which was irradiated to a lower fluence and annealed in shorter steps, it is not clear if it depends on the neutron fluence. Therefore, the module R9 needs to be further investigated with additional annealing steps.

For the module R3 further annealing steps will be performed to find the maximum efficiency of pixel design V5 and further research of the performance after reaching the maximum efficiency.

The higher hit detection efficiencies might be explained with charge multiplication which is observed in n-in-p micro-strip detectors after neutron irradiation and long term annealing \cite{Milovanoivi,Kramberger1}. To understand if the higher efficiency is caused by charge amplification, which was the intention of the REINER pixel designs, TCT measurements at different voltages and annealing steps are planned for irradiated modules. With this method the pixel designs can be investigated in the range of \textmu m to see which pixel parts might cause charge multiplication.

\acknowledgments
The authors would like to thank the team at the Sandia Annular Core Research Reactor, especially M. Hoeferkamp and S. Seidel, and the team at the TRIGA reactor in Ljubljana, especially V. Cindro, for their help with irradiation of the sensors.

Many thanks to all participants of the ATLAS ITk pixel test beam campaigns, especially those who develop and maintain the corresponding hardware and software.

The presented work is carried out within the framework of Forschungsschwerpunkt FSP 103 and supported by the Bundesministerium f\"ur Bildung und Forschung BMBF under grant 05H15PECA9.

This project has received funding from the European Union's Horizon 2020 Research and Innovation programme under Grant Agreement no. 654168.

The measurements leading to these results have been performed at the Test Beam Facility at DESY Hamburg (Germany), a member of the Helmholtz Association (HGF).

%\newpage



\begin{thebibliography}{99}

\bibitem{ATLAS}
ATLAS Collaboration, \emph{The ATLAS Experiment at the CERN Large Hadron Collider}, \emph{JINST} {\bf 3} (2008) S08003.

\bibitem{IBL}
ATLAS Collaboration, \emph{ATLAS Insertable B-Layer Technical Design Report}, CERN-LHCC-2010-013 (2010).

\bibitem{sensor1}
G. Aad, et.al., \emph{ATLAS pixel detector electronics and sensors}, \emph{JINST} {\bf 3} (2008) P07007.

\bibitem{sensor2}
S. Altenheiner, et.al., \emph{Planar slim-edge pixel sensors for the ATLAS upgrades}, \emph{JINST} {\bf 7} (2012) C02051.

\bibitem{FE-I4}
M.Garcia-Sciveres, et.al., \emph{The FE-I4 pixel readout integrated circuit}, \emph{Nucl. Instr. and Meth. A} {\bf A636} (2011) S155-S159.

\bibitem{REINER}
T. Wittig, \emph{Slim edge studies, design and quality control of planar ATLAS IBL pixel sensors}, \emph{Ph.D. Thesis} TU Dortmund (2013).

\bibitem{meinPaper}
M. Weers, et.al., \emph{Lab and test beam results of irradiated silicon sensors with modified {ATLAS} pixel implantations}, \emph{JINST} {\bf 13} (2018) C11004.

\bibitem{effVerlust1}
J. Weingarten, et.al.\emph{ Planar pixel sensors for the ATLAS upgrade: beam tests results}, \emph{JINST} {\bf 7} (2012) P10028.

\bibitem{effVerlust2}
ATLAS IBL collaboration, \emph{Prototype ATLAS IBL modules using the FE-I4A front-end readout chip}, \emph{JINST} {\bf 7} (2012) P11010.

\bibitem{Ljubljana}
K. Ambro\v zi\v c, et.al., \emph{Computational analysis of the dose rates at JSI TRIGA reactor irradiation facilities}, \emph{Appl. Radiat. Isot.} {\bf 130} (2017) 140-152.

\bibitem{DESYPaper}
R. Diener, et.al., \emph{The DESY II Test Beam Facility}, \emph{Nucl. Instr. and Meth. A} {\bf A922} (2019) 265-286.

\bibitem{resolution}
H. Jansen, et.al., \emph{Performance of the EUDET-type beam telescopes}, \emph{EPJ Tech. Instr.} {\bf 3} (2016) 7.

\bibitem{mimosa26}
J. Baudot, et.al., \emph{First test results of MIMOSA-26: A fast CMOS sensor with integrated zero suppression and digitized output}, \emph{IEEE} (2009) 1169-1173.

\bibitem{clopper-pearson}
C. J. Clopper, E. S. Pearson, \emph{The Use of Confidence or Fiducial Limits Illustrated in the Case of the Binomial}, \emph{Biometrika} {\bf 26} (1934) 404-413.

\bibitem{Andreas}
A. Gisen, \emph{Quad module prototypes and design improvement studies of planar n$^+$-in-n silicon pixel sensors for the ATLAS Inner Tracker upgrade}, \emph{Ph.D. Thesis} TU Dortmund (2018).

\bibitem{Moll}
M. Moll, \emph{Radiation damage in silicon particle detectors: Microscopic defects and macroscopic properties}, \emph{Ph.D. Thesis} U Hamburg (1999).

\bibitem{Milovanoivi}
M. Milovanovi, \emph{Effects of accelerated long term annealing in highly irradiated n$^+$-p strip detector examined by Edge-{TCT}}, \emph{JINST} {\bf 7} (2012) P06007.

\bibitem{Kramberger1}
G. Kramberger, \emph{Modeling of electric field in silicon micro-strip detectors irradiated with neutrons and pions}, \emph{JINST} {\bf 9} (2014) P10016.



%\bibitem{a}
%Author, \emph{Title}, \emph{J. Abbrev.} {\bf vol} (year) pg.

%\bibitem{b}
%Author, \emph{Title},
%arxiv:1234.5678.

%\bibitem{c}
%Author, \emph{Title},
%Publisher (year).


% Please avoid comments such as "For a review'', "For some examples",
% "and references therein" or move them in the text. In general,
% please leave only references in the bibliography and move all
% accessory text in footnotes.

% Also, please have only one work for each \bibitem.


\end{thebibliography}
\end{document}